%% file: main.tex
\documentclass[sigconf]{acmart}

\input{preamble/preamble}

\input{preamble/commands}

\input{preamble/setup}

\usepackage{textcomp}

\def\review{1}

\begin{document}

\title{The Show Must Go On - Examination During a Pandemic}
\author{Pamela~Fleischmann}
\affiliation{%
    \institution{Department of Computer Science, Kiel University}
    \city{Kiel} 
    \state{Germany}
}
\email{fpa@informatik.uni-kiel.de}

\author{Mitja~Kulczynski}
\affiliation{%
    \institution{Department of Computer Science, Kiel University}
    \city{Kiel} 
    \state{Germany}
}
\email{mku@informatik.uni-kiel.de}

\author{Dirk~Nowotka}
\affiliation{%
    \institution{Department of Computer Science, Kiel University}
    \city{Kiel} 
    \state{Germany}
}
\email{dn@informatik.uni-kiel.de}

\begin{abstract}
  \input{abstract}
\end{abstract}



\maketitle

\section{Introduction}
\input{introduction}

\section{Teaching}\label{secTeach}
\input{teaching}

\section{Examination}
\label{sec:exami}
\input{exam}

\section{Using our framework}
\label{sec:using}
\input{using}

\begin{figure}[t]
    \includegraphics[width=.9\columnwidth]{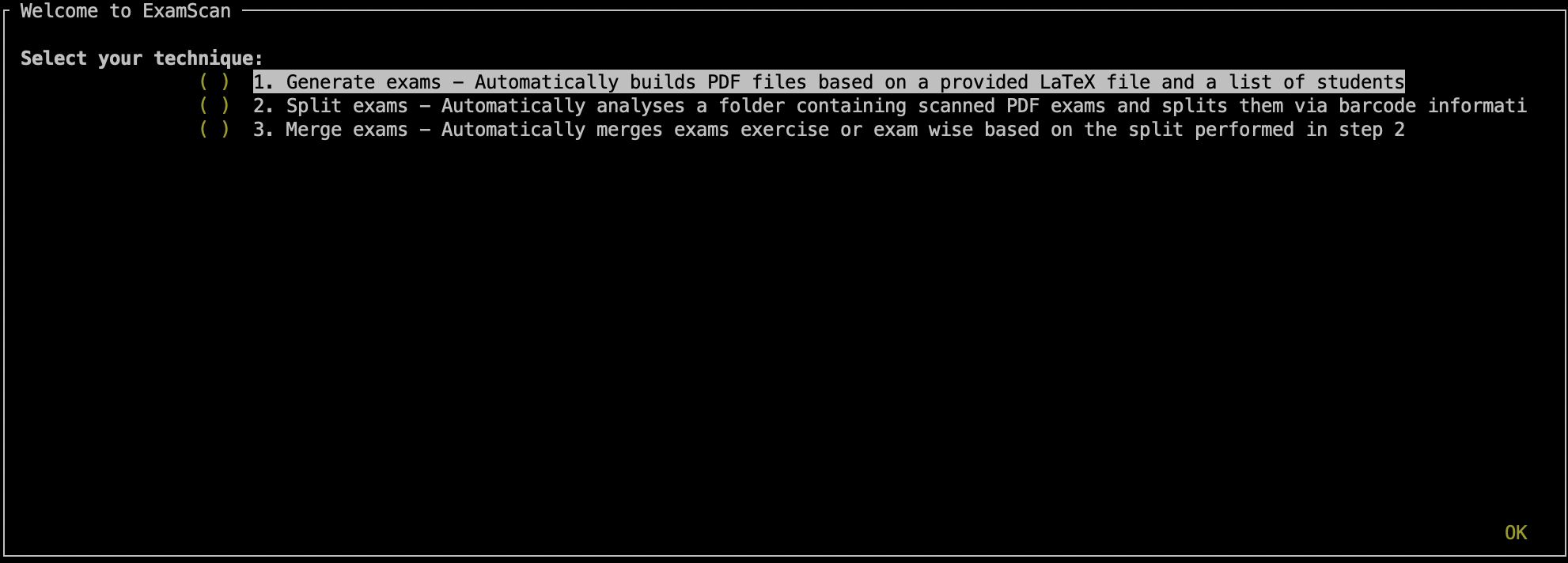}
    \caption{Terminal GUI for applying our algorithms\label{fig:gui}}
\end{figure}

\section{Conclusion}\label{sec:conc}
\input{conclusion}

\bibliographystyle{ACM-Reference-Format}
\bibliography{bibliography}

\end{document}

%% file: preamble/preamble.tex
\usepackage{paralist}

\usepackage{doi}
\usepackage[inline,draft]{fixme}

\usepackage[utf8]{inputenc}
\usepackage{xcolor}
\usepackage{syntax}
\usepackage{amsmath}
\usepackage{mathpartir}
\usepackage{wrapfig}
\usepackage{braket}
\usepackage{fontawesome}
\usepackage{listings}
\usepackage{todonotes}

\usepackage{hyperref}

\usepackage{tikz}
\usepackage{fontawesome}
\usetikzlibrary{positioning, arrows,automata,calc,patterns,shapes,shadows}
\tikzset{%
	unit/.style={draw=black,line width=0.15pt,rectangle, rounded corners=3pt,line width=0.15pt, color=black,fill=black!10,font=\tiny,text width=1.5cm, align=center, anchor=west,rounded corners=1pt},
	code/.style={text width=2cm,align=center,font=\huge}
}

\usepackage{afterpage}
\usepackage{pgfplots}

%% file: preamble/commands.tex
\colorlet{keywords}{green!40!black}
\colorlet{identifiers}{blue}
\colorlet{comments}{purple!40!black}
\colorlet{typecol}{green!40!black}

\newcommand{\cout}[1]{\ensuremath{c!}}
\newcommand{\cin}[1]{\ensuremath{c?}}

\usepackage{stackengine}

\definecolor{keywords}{HTML}{f9005a}
\definecolor{identifiers}{HTML}{333333}
\definecolor{comments}{HTML}{75715E}
\definecolor{typecol}{HTML}{679c00}

\lstdefinestyle{custompython}{
  belowcaptionskip=0pt,
  breaklines=true,
  frame=none,
  xleftmargin=1em,
  xrightmargin=-0.5cm,
  numbersep=2pt,
 language=python,
  showstringspaces=false,
  basicstyle=\scriptsize\ttfamily,
  keywordstyle=\bfseries\color{keywords},
  commentstyle=\itshape\color{comments},
  identifierstyle=\color{identifiers},
  stringstyle=\color{orange},
  numberstyle=\tiny\color{gray},
  numbers=left, 
  breaklines=true,
  postbreak=\mbox{{$\hookrightarrow$}\space},
}

\newcommand{\pythonstyle}[2]{\lstset{
  belowcaptionskip=0pt,
  breaklines=true,
  frame=none,
  xleftmargin=1em,
  xrightmargin=-0.5cm,
  numbersep=2pt,
 language=python,
  showstringspaces=false,
  basicstyle=\scriptsize\ttfamily,
  keywordstyle=\bfseries\color{keywords},
  commentstyle=\itshape\color{comments},
  identifierstyle=\color{identifiers},
  stringstyle=\color{orange},
  numberstyle=\tiny\color{gray},
  numbers=left, 
  breaklines=true,
  postbreak=\mbox{{$\hookrightarrow$}\space},
}}

\lstnewenvironment{pythonenv}[2]{\pythonstyle{#1}{#2}}{}

\newcommand\pythoninline[1]{{\pythonstyle{0}{}\lstinline!#1!}}

\newcommand{\univ}{\ifx\review\undefined our university\else Kiel University\fi}
\newcommand{\classlink}{\ifx\review\undefined\url{https://figshare.com/s/14135cce1c9762c6cf6d}\else\url{https://git.zs.informatik.uni-kiel.de/mku/latex_aufgabentemplate}\fi}
\newcommand{\frameworklink}{\ifx\review\undefined Conveniently we created a \textbf{reproduction package} tested within the virtual machine for VirtualBox running Ubuntu 20.04 LTS provided for the TACAS '21 artificial evaluation by Sebastian Hjort Hyberts, Peter Gjøl Jensen and Thomas Neele available at \url{https://zenodo.org/record/4041464\#.X\_hgBi1Q1QI}. Download the archive \url{https://figshare.com/s/df5854c5e8bd773b2161} and open a terminal within your virtual environment and extract the archive by executing \texttt{tar xf Downloads/repro.tar.gz -C /home/tacas21/}. The \texttt{readme.html} guides you through the installation and reproduction steps.\else\url{https://git.zs.informatik.uni-kiel.de/mku/examscan}\fi}
\newcommand{\coronalink}{\ifx\review\undefined\footnote{The link to our university specific COVID-19 regulations is omitted to not reveal our identities. In case of acceptance of our work, the link will be added.}\else~\cite{CoronaRegeln}\fi}

%% file: preamble/setup.tex
\AtBeginDocument{%
  \providecommand\BibTeX{{%
    \normalfont B\kern-0.5em{\scshape i\kern-0.25em b}\kern-0.8em\TeX}}}

\setcopyright{acmcopyright}
\copyrightyear{2018}
\acmYear{2018}
\acmDOI{10.1145/1122445.1122456}

\acmConference[Woodstock '18]{Woodstock '18: ACM Symposium on Neural
  Gaze Detection}{June 03--05, 2018}{Woodstock, NY}
\acmBooktitle{Woodstock '18: ACM Symposium on Neural Gaze Detection,
  June 03--05, 2018, Woodstock, NY}
\acmPrice{15.00}
\acmISBN{978-1-4503-XXXX-X/18/06}

%% file: abstract.tex
When unexpected incidents occur, new innovative and flexible solutions are required. If this event 
is something such radical and dramatic like the COVID-19 pandemic, these solutions must aim to 
guarantee as much normality as possible while protecting lives. After a moment of shock our 
university decided that the students have to be able to pursue their studies for guaranteeing a 
degree in the expected time since most of them faced immediate financial problems due to the loss 
of their student jobs. This implied, for us as teachers, that we had to reorganise not only the 
teaching methods from nearly one day to the next, but we also had to come up with an adjusted way 
of examinations which had to take place in person with pen and paper under strict hygiene rules.
On the other hand the correction should avoid personal contacts. We 
developed a framework which allowed us to correct the digitalised exams safely at home while 
providing the high standards given by the general  data protection regulation of our country. 
Moreover, the time spent in the offices could be reduced to a minimum thanks to automatically 
generated exam sheets, automatically re-digitalised and sorted worked-on exams.

%% file: introduction.tex
At the mid of March, \univ{}
decided that the summer semester, starting on April 
1$^\mathrm{st}$, has to be held as normal as possible, not in person but  via online 
teaching until the pandemic situation becomes clearer. Since online teaching was despite some 
platforms 
for uploading code, announcements, and distributing exercise sheets as well as marking the 
solutions, not established for teaching in the 
Department of Computer Science. Solutions were needed and implemented within two weeks. 
Since we are responsible for teaching {\em Theoretical Foundations of Computer Science (TFoCS)}
in the bachelor's programme at the Department of Computer Science of \univ{} during summer 
semesters, we were faced with the situation to teach one of the hardest courses in the bachelor's 
programme in a completely new way for approximately 250 students. From March to May, reliable 
information from \univ{} 
were understandably only provided for the following week and thus long-term planning was 
impossible, especially for the examination phase in summer as well for the postponed second 
examination phase (normally at the end of March) from the previous winter semester. This absence of 
planning dependability clouded the complete semester for us as teachers as well as for the students 
whose uncertainty of the situation had to be covered in the tutorials next to their struggling with 
the content. In May it became clear that the rest of the semester will take place online but that 
the examinations have to be in person as long as the pandemic situation changes for the better. 
Thus, we knew that we are provided with a 
lecture hall in which the students are able to write an exam with at least 2 m distance between one 
another. Additionally, we got hygiene rules which had to be followed before, during, and after the 
examination which we had to care of. On one hand, we were relieved that the exam would take 
place nearly normally, but on the other hand it was clear that the correction of the paper-written 
exams could not take place as usual. Normally, one team member corrects one exercise of all exams 
and thus we meet with all team members in one room and pass the exams to the next corrector once 
an exercise of one exam is corrected. The 
conception that each of us corrects in their own office (notice that most of us have not been in 
their offices for months) and we regularly exchange stacks of exams with 2 m distance, did not 
convince us. This 
procedure would also have included that we wear rubber gloves for the whole time or we vent the 
exams for two hours between passing. Thus, the idea of digitalising the exams emerged: scanning the 
hand-written exams, correcting them from home, and uploading the corrected versions for the 
students to review. Consequently, a system was required such that the digital 
versions of the exercises  can be mapped uniquely to the students while still being related to a 
single 
exams. Such a system allows us to organise the exams exercise-wise for correction and 
rearranging them student-wise for the review process by the students. Notice that by the 
digitalisation of the personalised exams, we had to fulfil the {\em general data 
protection regulation} (GDPR) \cite{dsgvo}.
The need of an own system arose after searching the literature. Preliminary experiments 
with similar approaches to digital postprocessing of exams (cf. \cite{pdfdicer})
turned out to be not as reliable in our setting. Second, the proposed schedule of other
approaches is not build for an exam-like setting. Therefore, as a consequence and in 
the need to
deal with our and most likely other university's situations (paper-written exam in person while 
meeting the Covid19 regulation), we came up with an own, highly 
adaptable framework. To keep up our 
schedule we developed
an extensible framework allowing us to automatically cope with this processes online to avoid 
contact 
in person as much as possible. To this extend we share not only the framework but also the insights 
on 
how we came up with the ideas. Secondly, we developed a \LaTeX{} class providing an easy to use tool 
box for creating exams and
exercise sheets.

The paper is structured as follows: In Section~\ref{secTeach}, we shortly describe how teaching is 
normally performed (without a 
pandemic, home office, and online teaching), and then we elaborate how we managed to teach the 
lecture (TFoCS) and the associated tutorials in 
this very special semester. Afterwards, in Section~\ref{sec:exami} we explain in detail how the 
examinations were conducted including the examination itself as well as the correcting part and 
the ensuing return including the discussion with single students afterwards.  In 
Section~\ref{sec:using}
we wrap things up and give a short introduction in how to configure the tool. In the last session 
we summarise our experiences and compare it to other approaches conducted in comparable courses at 
\univ{}.

%% file: teaching.tex
Normally, the course TFoCS consists of a lecture and associated tutorials which each take place 
every 
week. Each week the students get 
a sheet of exercises which they can hand-in after one week for corrections. The format of the 
lecture is classically teacher-centred, where the lecturer presents the contents on a white- 
or blackboard for twice one-and-a-half-hour per week. For the tutorials, we 
have two forms: supervised learning time (SLT) and a teacher-centred tutorial (TCT). The SLT is 
offered twice a week for each four hours. During this time the students and a teacher meet in a room 
and the students work on the homework or the content of the lectures on their own but have a 
professional person 
available for asking questions. The TCT takes place once a week for one-and-a-half hour and in a 
dialogue, exercises similar to the homework (or the solutions of already corrected homework) are 
discussed. Both the lecture and the tutorial take place in person and are not recorded. Moreover, 
most of the students hand in their homework handwritten on paper, and the corrections are as well on 
the paper and handed back in the following tutorial. Only for announcements regarding the lecture or 
the tutorial, for uploading the exercise sheets as well as the lecture notes, and for managing the 
obtained points for the homework, the \textsc{OpenOlat}-platform \cite{olat}, hosted by \univ, is 
used. Attempts from our side to establish more online learning, e.g. a forum in \textsc{OpenOlat} for 
discussions, were not used by the students in the past years. This form of teaching was managed by 
one professor (for the 
lecture), three PhD-students for the tutorials, and three student assistants, hired for 50 
hours per week, for the corrections.

\noindent\paragraph{Adjusting to Online Teaching}
Within in the two weeks of preparing the online semester and the discussion how we can guarantee 
that the students are able to get in touch with us and with each other, the data centre of \univ{}
established and provided BigBlueButton \cite{bbb} as a video conference platform and the 
Department of Computer Science ordered IPads by Apple\texttrademark. Thus, we decided that large 
parts of the teaching and especially the interaction with the students, take place in 
BigBlueButton rooms. Nevertheless, we agreed that a lecture with 250 students via BigBlueButton 
would not be the same as in a lecture hall and thus, we opted to produce short videos (ten to 15 
minutes) explaining the week's content, to upload them on the \textsc{OpenOlat}-platform, and to offer the 
tutorials to discuss the content. Therefore, the tutorial time was restructured as well: fourth a 
week we offered tutorials for one-and-a-half hour via BigBlueButton where exercises covering the 
week's content were presented and worked on interactively with the students. For getting smaller 
groups of students, at two time slot two parallel groups were offered.  Moreover, for being 
responsive for the very different needs of the students (cf. \cite{fleischmann2019managing}), one 
time slot was dedicated for beginners, one for advanced students, one for the students aiming 
for a bachelor in education and one in English language. Since we did not want to lose the idea of 
SLT completely, we additionally offered a time slot where the students could ask specific questions 
regarding the content of the lecture or the homeworks. Thus, we extended the TCT  to 9 hours and 
reduced the SLT  by 6.5 hours in comparison to the last years because we did not believe that the 
SLT where students work in small groups and only reach out for the supervisors when they have 
questions is easily transformable into a BigBlueButton conference. Since more tutorials took place 
in parallel, the team enlarged by one PhD-student, one student assistant hired for 50 hours per 
week, and one student assistant hired for 25 hours per week. The small videos were completely 
produced by the professor, while the TCT and SLT was covered by the PhD students. The student 
assistants were responsible for the correction of the homeworks. These corrections were in more 
detail than in the previous years to expound the problems and the very points where something went 
wrong since we believed that they need more guidance throughout this special semester.

\noindent\paragraph{Technical Summary.} Exchanging document with the students as well as notifying students 
about changes to the weekly routine was managed with the \textsc{OpenOlat} platform. Moreover we used 
\textsc{OpenOlat} for grading the students. For the interaction with students we used BigBlueButton for 
groups of 
students and mainly email for the communication with single students; in rare cases we also used 
common chat clients for a synchronous communication. It is worth remarking that for the tutorials 
in BigBlueButton we attached a tablet and used the \textsc{Notability}-App \cite{notability} to 
simulate the whiteboard from a classroom.

\noindent\paragraph{Feedback.}
After half a semester, we conducted a midterm evaluation to get the possibility to adjust the new 
and unexperienced way of teaching to the students' needs. Fortunately, the majority of the students 
approved of the way we managed teaching. Nevertheless a lot of students missed the help and the 
meetings in persons - not only with us but also with each other for working on the exercises.
For us astonishingly, most of the students neither used the camera nor the microphone during the 
tutorials for the reason that they either do not have a camera or a microphone or were afraid of 
disturbing the flow of the tutorial. The evaluation at the end of the semester does not differ much 
from the midterm evaluation even though we encouraged the students to use the microphone to ask 
questions instead of writing their questions (mostly formulae) in the BigBlueButton chat. In 
general, the students 
liked the extended offers for help but missed 
the option to ask questions in person since it is harder to formulate questions in a chat than in a 
personal conversation. Moreover, we got more complaints about BigBlueButton, namely overloads and 
not adjustable audio.

\noindent\paragraph{Our Impression.}
After a rough start, we equipoised to the technique and the used platforms. Not surprisingly, the 
most important part for online teaching is a working technique, reliable software, and 
flexibility from all participating groups. The first two parts worked unfortunately only more or 
less during the semester: bad quality in the online tutorials via BigBlueButton as well as not 
reachable platforms due to too many persons accessing them at the same time, made it sometimes hard 
to provide good teaching. Most of the participants - students as well as teachers - showed a huge 
amount of flexibility without complaining. From our point of view the students had more problems to 
adjust to the new situation than the teachers for two main reasons: it is common that after a 
lecture or a tutorial some students reach out to the teacher to ask questions in private. Even 
though we stayed in the BigBlueButton conference after the event ended, nobody stayed as well to 
ask questions. We believe that the students who do not dare ask questions while other students 
listen, were also afraid that somebody is still in the conference who may blame them for their 
questions. This impression is tightly related to the second point: the students have problems to 
formulate their questions. Since we do not believe that there is any relation between formulating 
questions and the pandemic, we had to learn that this problem probably exists for a longer time.
Normally, the students meet us directly after the tutorial or lecture, or they come to our offices, 
and show us what they did on a sheet of paper. In a direct conversation they are able to point at 
problematic parts they have written. Since most of our students did not dare turn on their 
camera or even the microphone, they had to write all questions which is more complicated than 
pointing with the finger on something. We encouraged them to take photos and upload them but we 
noticed that this solution is not working for them as well. We were able to capture some of these 
students via common chat clients where they felt safer to ask questions which were not formulated 
in detail. During nearly the complete semester the TCT for beginners and advanced students was 
frequented by 30 students in average, the tutorial in English language by 15 students in average, 
and the tutorial for the students aiming for a bachelor in education by six students in average.
Notice that the students were allowed to participate in each tutorial they liked, even more than 
one per week.
For us interestingly, the students of the beginner groups were more active in the sense by 
participating in polls (provided by BigBlueButton), answering question, and asking questions on 
their own even beyond the scope of the week's content. As we assumed the SLT was rarely 
frequented and mostly by students who wanted us to work on their homework step by step with 
them which is not the idea of SLT. Specific for this semester during the 
pandemic and not an online semester in general was 
in addition that the students needed more care despite the professional part: they were scared by 
the pandemic and the resulting uncertainty of their exams.

%% file: exam.tex
At \univ{} each undergraduate course has a final written exam in person to pass the course. 
This exam can be 
taken in one of two exam's periods: one  at the end of the semester and  one at the beginning of 
the next semester. The exam consists of a multiple choice section (reproduction of knowledge) and 
seven further problems, each of them similar to, but in general easier than the problems 
of
the homework exercises seen during the semester. The exam is set up for three hours. Supervised by 
the 
teaching assistant the exams are corrected by the team members together in 
a meeting room. This process usually takes three working days and requires working closely together. 
The final part of an exam is a review process of the corrections by the students. During an hour 
period the students have the chance to scrutinise their corrected exercises and mistakes in the 
correction from our side can be straightened. This step is also held 
in person in a meeting room holding roughly ten persons.

Since the Department of Computer Science disallowed online exams, we established a 
solution accommodating the formal regulations made by the head of the 
university\coronalink. The exam itself was finally taken in a marquee having 1000 $m^2$ 
accommodating 300 students. To keep the correction process as usual as possible and still 
avoid contact in person we developed an own, nearly automatic solution to digitalise the 
exams, correct the digital version, and provide an online review of the digitally corrected 
exams.
\begin{figure}[t]
    \includegraphics[width=.9\columnwidth]{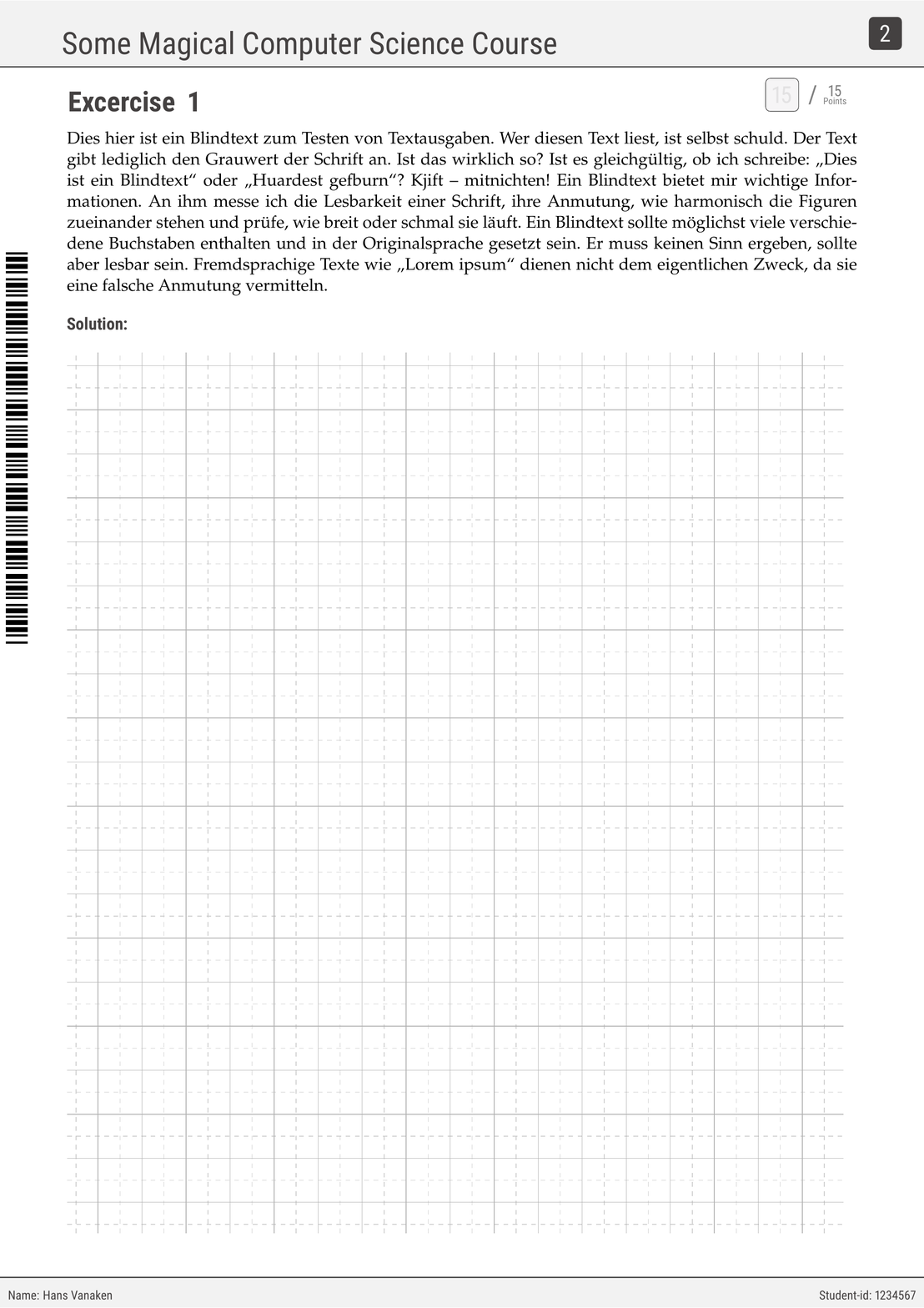}
    \caption{An exemplary exam page including a barcode\label{fig:exam}}
\end{figure}

Letting a computer do the  work for us required the implementation of an appropriate format. Each 
exam was already 
personalised to a particular student having their name and their student identification number on 
each page in the 
past years. However, 
reading plain text from scanned sheets of paper is not an easy task for a computer. Therefore, we 
decided to generate 
barcodes on every exam page as depicted in \autoref{fig:exam}. Our choice of an Code-39 barcode was 
made after performing empirical experiments (scanning crumbled paper, slightly folded paper, as 
well as partially hand-writings on the code) with various encoding format e.g. QR-codes. Since our 
exams are created and compiled in \LaTeX{} we are able to use the CTAN package 
\textsc{makebarcode}~\cite{makebarcode} for generating barcodes encoding the student 
identification number, the page number, and 
the exercise number. Therefore, it is an easy task for a computer to gather the required unique 
information of a single page. By the additional information (page number and exercise number) in 
comparison to the previous years, we were also unburdened of the task to order the single pages 
of the exam by hand. In \autoref{lis:latex} we show the import of the CTAN package and an 
exemplary invoking of a barcode.

\begin{lstlisting}[mathescape=true,style=custompython,caption=Invoking barcodes in 
\LaTeX{},label=lis:latex]
# import the makebarcode package
\usepackage[code=Code39,X=.5mm,ratio=2.25,H=0.5cm]{makebarcode}

# Generating the barcode
\barcode{\StudentID-\arabic{ExerciseNo}-\thepage}
\end{lstlisting}
\noindent
The generation of the exams is performed by using a \texttt{csv}-file containing the students 
credentials. 
Our \LaTeX{} basis file is enriched by macros being replaced during the generation of each 
personalised exam. Finally, the individual \LaTeX{} source is translated by using 
\textsc{latexmk}~\cite{latexmk}. Thus, by the interaction of the exam's source code and the list of 
students in a \texttt{csv}-file, all personalised exams are generated in one step.
We outline the algorithm in \autoref{lis:generate}.

\begin{lstlisting}[mathescape=true,style=custompython,caption=Generate exams with barcodes 
automatically,label=lis:generate]
with open('participants.csv') as student:
  reader = csv.DictReader(student,fieldnames=self.__sdata["fieldnames"]
    for row in reader:
      exam = exam_tex_file.tex
      generated_exam = f"{outputFolder}/{row[self.__sdata["key"]]}.pdf"
      for line in exam:
        for fieldName in fieldNames:
          line = line.replace(fieldName,row[fieldName])
          generated.write(line)
      run_latexmk(generated_exam)
\end{lstlisting}
\noindent
Before the exam, each student got their own personalised printed version in an envelope. During 
the exam, the exercises are solved using a pen as in the previous years. Afterwards, we got the 
exams back bagged
again in the envelopes. Then, we scanned all exams into a single, large \texttt{PDF}-file. Since 
the pages were not necessarily sorted, 
flipped upside down, or rotated  (we decided not to staple each exam such that we do not have to 
remove the retaining clips for scanning) our automatic processing had 
to take care of this step as well. Scanning of many pages at the same time naturally causes a 
slightly rotation on some pages and therefore noise within the received data. The decoding process 
is manifold: firstly, using \textsc{ghostscript}~\cite{ghostscript}, each pages was converted in to 
a \texttt{JPEG}-picture interpolated by using Bresenham's line algorithm~\cite{5388473} throughout the entire 
picture. This 
algorithm plots the closest texture coordinate at each pixel resulting in a much better readability 
of our barcode -- especially for a computer. Second, we extracted the particular part of the page 
where the barcode is rendered. This step was again performed by invoking \textsc{ghostscript}. The actual 
decoding of the image was performed by \textsc{ZBar}~\cite{zbar}, an open source tool supporting 
Code-39 
barcodes. 
We extracted each page, and renamed it to the decoded string of student identification number, page 
number, and exercise number as shown in \autoref{lis:splitpages}.

\begin{lstlisting}[mathescape=true,style=custompython,caption=Split the scanned exams into single 
pages,label=lis:splitpages]
for page in scanned_pdf:
  code = decode_barcode(page) # invoke zbar-img to decode the barcode
  create_folder(student_id)
  extract_pdf(page,f"{student_id}/{code}.pdf") # invoke ghostscript to extract a single PDF page
\end{lstlisting}

The way we invoked the previously described techniques turned out to be extremely reliable. Within 
our 
two exams each consisting of more than 3000 pages, we only failed to decode 10 pages. These errors 
where caused by too heavy rotations of the scanned pages or handwritings causing too much noise on 
the barcode.

\smallskip

To ease the manual correction procedure, the pages can be merged into either exercise-wise 
files or student-wise files. As mentioned before, since in our scenario one team member corrects a 
single exercise of all exams, the first choice is consequently the natural one for correction. By 
using the data gathered by decoding 
the barcodes, we sort all pages ascending to their relative page number in each exam and then merge 
all files having the same exercise number. The algorithm is outlined in \autoref{lis:merge}.

\begin{lstlisting}[mathescape=true,style=custompython,caption=Merge single pages for further 
processing,label=lis:merge]
for student_id_folder in processed_folder:
  if not split_exercise: # merge all documents of a student in the correct order
    merge_pdfs(all_pages(student_id_folder),f"{outputfolder}/{student_id}.pdf")
  else: # merge all exercises in different PDF files
    create_folder(f"{outputfolder}/{student_id}/")
    collected_file_names = dict()
    for file_name in list_files(student_id_folder): # collect files names for an exercise
      student_id,page_no,exercise_no = file_name.split("-")
      if not exercise_no in collected_file_names:
        collected_file_names[exercise_no] = []
      collected_file_names[exercise_no]+=[file_name]
      for exercise_no in collected_file_names: # merge PDFs and store them separately
        
merge_pdfs(collected_file_names[exercise_no],f"{outputfolder}/{student_id}/{exercise_no}.pdf")
\end{lstlisting}

\begin{figure}[t]
    \includegraphics[width=.9\columnwidth]{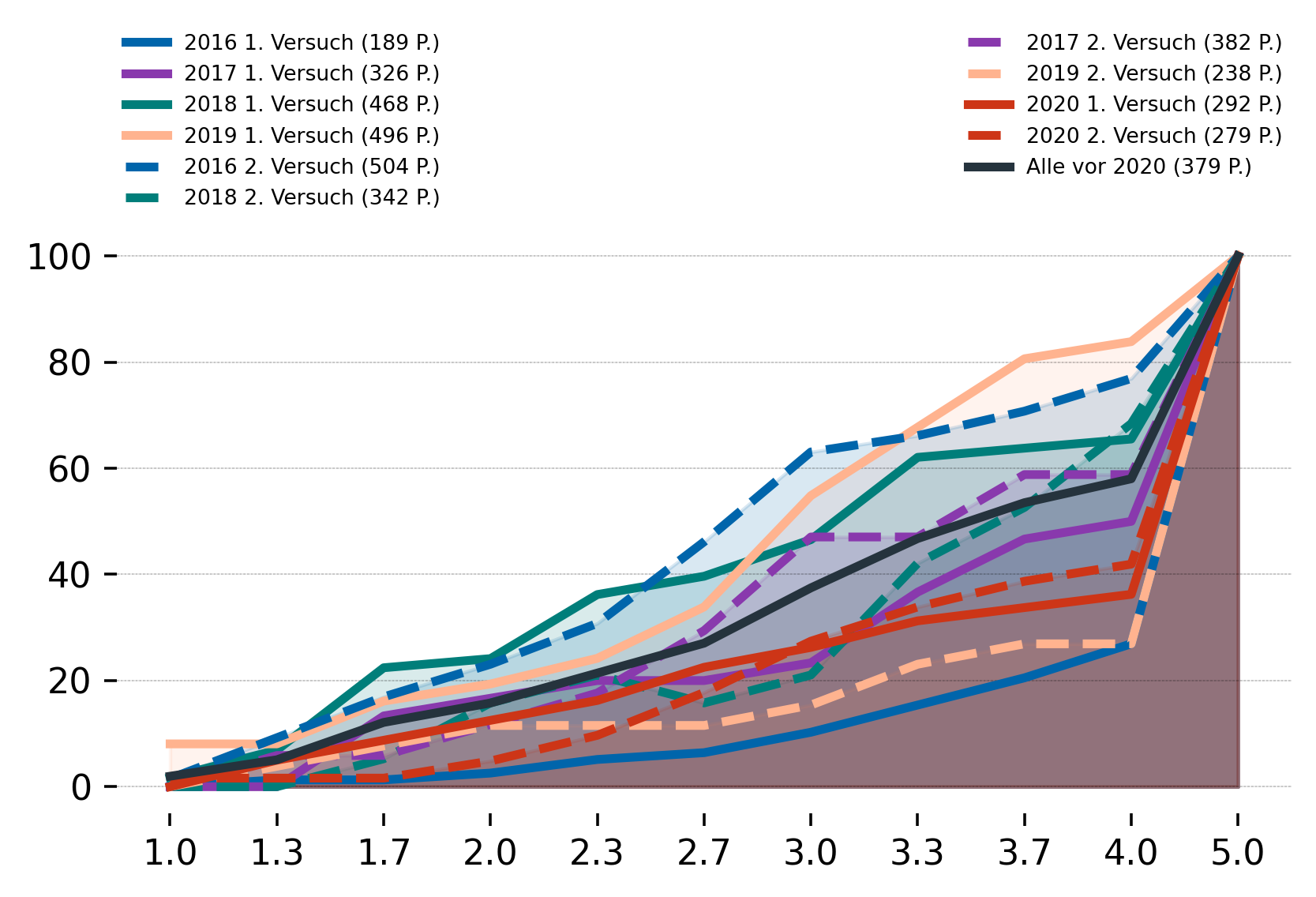}
    \caption{Exemplary output of the gathered exam data during corrections\label{fig:grades}}
\end{figure}

Afterwards the resulting files were stored in a private \textsc{NextCloud}~\cite{nextcloud} allowing an easy 
access and 
modification by multiple team members while meeting the GDPR. For the correction we use again the 
\textsc{Notability}-App~\cite{notability} 
which offers a build in link to WebDAV services being served by our \textsc{NextCloud} server. Each team 
member corrected a single exercise student by student and pushed 
the results back into the \textsc{NextCloud}. Again, this synchronisation is performed fully automatically. 
Achieved points were automatically collected into a \texttt{csv}-file allowing easy post processing 
of the 
data -- e.g. plotting grade distributions as shown in \autoref{fig:grades}.

Since the Department of Computer Science guarantees the students a review of the marking,
we had to make the corrected exams available for the students. Thus, for this purpose the 
corrected single pages of the exams were sorted student-wise: all exercises of each 
student were merged back into a single file and uploaded to \textsc{OpenOlat}. Each student is now 
able to review their own exam including the corrections made by the team members. For the 
discussion we offered a BigBlueButton conference where the team members could answer individual 
questions in break-out rooms in a private communication while sharing the digital and corrected
version of the specific exam via BigBlueButton.

%% file: using.tex
We believe that our framework is benificial for many other similar settings other than the one we face at \univ. We therefore decided to build a flexible and easy to reuse tool serving all benefits of our described approach. 


\paragraph{Creating the exam}
To ease the initial step of creating the exam we provide a \LaTeX{} class\footnote{The \LaTeX{} 
class is available at \classlink}. This 
class contains many useful features e.g. an environment for solutions, changing colours to a 
printer friendly schema, or an exam title page holding an overview of exercises and associated 
points, as well as the invoking of barcodes. The important part is the former mentioned generation 
of barcodes on each page which is performed automatically by adding \texttt{barcode} as an option to 
our document class. The following line loads the \LaTeX{} class and activates, next to barcodes, an exam
title page, activates printer friendly mode, and disables the solution environment.
\begin{center}
\pythoninline{\\documentclass[hidesolution,bw,barcode,titlepage]\{aufgabe\}}
\end{center}
A full documentation of implemented class options is available within the read-me file of our \LaTeX{} class.

\paragraph{Creating the students list}
The list of students has to be stored in a \texttt{csv}-file containing the last name, the first 
name, the student identification number, and the email-address. Thus, an exemplary row in the file 
has the following form
\begin{center} \pythoninline{Vanaken;Hans;372048;hans.vanaken@some-uni.eu}\end{center}
Notice that the {\em columns} are separated by semicolons. 

\paragraph{Setting up the framework}The presented framework\footnote{\frameworklink} -- purely written in \textsc{Python 3} -- 
requires three external tools installed within your system's paths:  
\textsc{ghostscript}~\cite{ghostscript}, \textsc{ZBar}~\cite{zbar}, and 
\textsc{latexmk}~\cite{latexmk}. The executables are linked within a JSON file named 
\pythoninline{toolconfig.json} being preconfigured to cope with Apples OSX and Ubuntu Linux. Notice 
that these executable paths have to be adapted whenever using a different operating system.

The linking of the students data within the \texttt{csv}-file is configured within the JSON file \pythoninline{student_data.json}. We 
define the path to our \texttt{csv}-file holding the students data, a list of column identifiers 
naming each 
entry inside the \texttt{csv}-file, and a key for each row, e.g. StudentID for the student 
identification number as exemplarily shown in 
\autoref{lis:studentcsv}.

\begin{lstlisting}[mathescape=true,style=custompython,caption=Student data setup file \texttt{student_data.json},label=lis:studentcsv]
{
  "student_data" : {
    "file_path" : "./participants.csv",
    "fieldnames" : ["LastName","FirstName","StudentID","Email"],
    "key" : "StudentID"
  }
}
\end{lstlisting}
The \texttt{csv}-file can be enriched by all information necessary for conducting an exam in a 
specific university. Notice that the \pythoninline{student_data.json} has to be adjusted 
accordingly and that renaming the {\em fieldnames} (cf. \autoref{lis:studentcsv}) requires an 
analogous renaming of the macros placed within the \LaTeX{} exam file (e.g. replacing \texttt{\#\#FIRSTNAME\#\#} with \texttt{\#\#VOORNAAM\#\#} with the intention of changing macros to Dutch).


\paragraph{Executing the algorithms}
After finishing the setup one calls the \textsc{Python 3} script \pythoninline{examScan.py} by simply executing 
	\pythoninline{python3 examScan.py}
This call conveniently opens an elementary Graphical User Interface (GUI) being depicted in 
\autoref{fig:gui}. The GUI now allows selecting between our algorithms described in 
Section~\ref{sec:exami} by using your keyboard's arrow keys.

Summing the necessary steps up, the following parts are required to perform an exam with our 
framework:
\begin{enumerate}[1.]
\item Create an exam with \LaTeX{} by using our provided template\footnote{The template is available at \classlink. It contains next to examples a read-me file guiding you through the setup.}.
\item Create a students list as \texttt{csv}-file.
\item Generate and compile the personalised exams into one \texttt{pdf}-file by using the 
framework. Select technique \texttt{1} within the the GUI depicted in \autoref{fig:gui}).
\item Print the previously obtained \texttt{pdf}-file.
\item Scan the worked-on exam sheets. (Notice that you do not have to sort them in before).
\item Split the obtained \texttt{pdf}-file in one \texttt{pdf}-file per sheet by using the 
framework. Select technique \texttt{2} within the GUI.
\item Merge the obtained \texttt{pdf}-files exercise-wise or student-wise by using the framework. 
Select technique \texttt{3} within the GUI.
\item Distribute the obtained \texttt{pdf}-files for correction, meeting your specific regulation 
for 
data protection (in our case \textsc{NextCloud}).
\item Correct the exercises with annotations or by tablet (we used the \textsc{Notability}-App).
\item Make the exams (student-wise sorted) available for a review process by the students.
\end{enumerate}

%% file: conclusion.tex
Faced with a situation non of us expected, the discussion about online teaching formats and online 
examination speeded up. All arguments against online teaching became meaningless from one day to 
the next due to a lack of other options. The necessity of deciding fast about {\em 
what do we like in classical teaching and what to we want to keep?} as well as {\em which new 
possibilities do we have with online teaching?} brought a new way of teaching. Next to the 
teaching, we were confronted with the examination situation: all exams from the 
undergraduate courses highly depend on reproducing the 
content or solving small problems similar to the exercises of the homework. The solutions to these 
exercises can be easily found in the internet.
Discussions in the Department of Computer Science about changing the form of the exercises did not 
result in convincing alternatives. For 
this reason and a relaxed pandemic situation in the summer, the Department of Computer Science 
decided that the exams have to take place in person including strict regulations regarding the 
hygiene. Since, we wanted to minimise contacts during the correction part, we built a framework for 
a digital correction and an online review. With this framework, we were gained the following 
benefits
\begin{itemize}
\item automatically generated exams in one \texttt{pdf}-file,
\item automaticall sorted exams exercise-wise or student-wise for corrections or the review 
respectively,
\item relaxed corrections from home without meetings in person,
\item fast and efficient review process since the students were able to check the corrections 
beforehand.
\end{itemize}
The experiences we gathered showed that digitalising the exams and correcting with annotations or a 
tablet are not as difficult as expected once an appropriate framework is provided. Moreover, the 
advantages that first the exhausting work of correcting 150 times the same exercise is easier from 
home and second wrong corrections during the correction process can be set aright easily with 
annotations rather than with a pen. We noticed that we are even faster in correcting the exams in 
comparison to the previous years.

At our university other groups and faculties decided differently how to manage the examinations 
(for teaching it was required to switch to online teaching completely). We briefly compare their 
experiences with ours. The most remarkable situation, we know of, occurred in the Department of 
Economics: they decided to perform the written exams online with the result that instead of the 
usual 30\% of students passing this year 80\% passed, i.e. cheating could not be avoided. Other 
groups corrected in university, thus one single person per office. They faced the situation that 
the organisation of passing the exams among half a dozen correctors should not be underestimated 
for all having continuously exams to correct. Regarding the review process, some groups decided to 
meet in similar environments as for the exams: huge lecture hall, the student get the corrected 
exams at the door, take a seat, review the corrections, and afterwards may ask questions through a 
plexiglass window. This procedure had on the one hand the advantage that the students and 
correctors met in person and were able (with mask and rubber gloves) to point with the finger on 
unclear parts of the corrections. On the other hand, this procedure took several hours, since first 
the students had to keep distance while entering the room and obtaining their exams, second the 
students saw the corrections the first time at this point and they had to wait for a long time 
until they were allowed to ask questions. As in the exams themselves, an organised way of entering 
and leaving the lecture hall such that each two students never meet with a distance of under 2m and 
are registered with their arriving and leaving time for a later on traceability, takes 
approximately the tripled time to the situation from the previous years. With our solution, first 
the cheating potential was not higher than before since the exams took place in person and even 
with more distance. Second, the correction process was speeded up since by the automatically 
exercise-wise order of the exams, passing exams and waiting for colleagues was omitted. Third, and 
for us the most important part, is the advantage of the review process: since the students were 
able to see the digital corrections online before the actual review process, they could ask 
immediately specific questions. Therefore, the review process was faster and more efficient than in 
the previous years. Please notice, that due to the new and special situation the students did not 
have to sign the corrected exams as it is required normally. Nevertheless, even if the review 
process is going to take place in person again, we will keep on with the digitalisation of the 
exams, the digital corrections, and the provision of the digitally corrected exams since the time 
savings and the more relaxed corrections are convincing for us.

We highly believe -- not only due to frequent questions about our efficient way of coping with the ongoing situation within our
university and one other university within our country -- that our attempt forms a great start to making teaching and examination easier and less time consuming in the future by still keeping up
high standards.